\documentclass[8.5pt,twoside,twocolumn]{article}
\pdfoutput=1
\usepackage[super,sort&compress,comma]{natbib} 
\usepackage[version=3]{mhchem}
\usepackage{times,mathptmx}
\usepackage{sectsty}
\usepackage{balance} 
\usepackage{amssymb} 
\usepackage{color}

\usepackage{graphicx} %eps figures can be used instead
\usepackage{lastpage}
\usepackage[format=plain,justification=raggedright,singlelinecheck=false,font=small,labelfont=bf,labelsep=space]{caption} 
\usepackage{fancyhdr}
\begin{document}

\thispagestyle{plain}
\fancypagestyle{plain}{
%\fancyhead[L]{\includegraphics[height=8pt]{headers/LH}}
%\fancyhead[C]{\hspace{-1cm}\includegraphics[height=20pt]{headers/CH}}
%\fancyhead[R]{\includegraphics[height=10pt]{headers/RH}\vspace{-0.2cm}}
\renewcommand{\headrulewidth}{1pt}}
\renewcommand{\thefootnote}{\fnsymbol{footnote}}
\renewcommand\footnoterule{\vspace*{1pt}% 
\hrule width 3.4in height 0.4pt \vspace*{5pt}} 
\setcounter{secnumdepth}{5}

\makeatletter 
\def\subsubsection{\@startsection{subsubsection}{3}{10pt}{-1.25ex plus -1ex minus -.1ex}{0ex plus 0ex}{\normalsize\bf}} 
\def\paragraph{\@startsection{paragraph}{4}{10pt}{-1.25ex plus -1ex minus -.1ex}{0ex plus 0ex}{\normalsize\textit}} 
\renewcommand\@biblabel[1]{#1}            
\renewcommand\@makefntext[1]% 
{\noindent\makebox[0pt][r]{\@thefnmark\,}#1}
\makeatother 
\renewcommand{\figurename}{\small{Fig.}~}
\sectionfont{\large}
\subsectionfont{\normalsize} 

\fancyfoot{}
%\fancyfoot[LO,RE]{\vspace{-7pt}\includegraphics[height=9pt]{headers/LF}}
%\fancyfoot[CO]{\vspace{-7.2pt}\hspace{12.2cm}\includegraphics{headers/RF}}
%\fancyfoot[CE]{\vspace{-7.5pt}\hspace{-13.5cm}\includegraphics{headers/RF}}
\fancyfoot[RO]{\footnotesize{\sffamily{1--\pageref{LastPage} ~\textbar  \hspace{2pt}\thepage}}}
\fancyfoot[LE]{\footnotesize{\sffamily{\thepage~\textbar\hspace{0cm} 1--\pageref{LastPage}}}}
\fancyhead{}
\renewcommand{\headrulewidth}{1pt} 
\renewcommand{\footrulewidth}{1pt}
\setlength{\arrayrulewidth}{1pt}
\setlength{\columnsep}{6.5mm}
\setlength\bibsep{1pt}

\twocolumn[
  \begin{@twocolumnfalse}
\noindent\LARGE{\textbf{Universal ultracold collision rates for polar molecules of two alkali-metal atoms}}
\vspace{0.6cm}

\noindent\large{\textbf{Paul S. Julienne,$^{\ast}$\textit{$^{a}$} Thomas M. Hanna,\textit{$^{b}$} and
 Zbigniew Idziaszek\textit{$^{c}$}}}\vspace{0.5cm}
%Please note that \ast indicates the corresponding author(s) but no footnote text is required. 

\noindent\textit{\small{\textbf{
Submitted to Physical Chemistry Chemical Physics,  themed issue on cold molecules, 2011}}}
%Received Xth XXXXXXXXXX 20XX, Accepted Xth XXXXXXXXX 20XX\newline
%First published on the web Xth XXXXXXXXXX 200X}}}

%\noindent \textbf{\small{DOI: 10.1039/b000000x}}
\vspace{0.6cm}
%Please do not change this text.

\noindent \normalsize{Universal collision rate constants are calculated for ultracold collisions of two like bosonic or fermionic heteronuclear alkali-metal dimers involving the species Li, Na, K, Rb, or Cs.  Universal collisions are those for which the short range probability of a reactive or quenching collision is unity such that a collision removes a pair of molecules from the sample.  In this case, the collision rates are determined by universal quantum dynamics at very long range compared to the chemical bond length.  We calculate the universal rate constants for reaction of the reactive dimers in their ground vibrational state $v=0$ and for vibrational quenching of non-reactive dimers with $v \ge 1$.  Using the known dipole moments and estimated van der Waals coefficients of each species, we calculate electric field dependent loss rate constants for collisions of molecules tightly confined to  quasi-two-dimensional geometry by a one-dimensional optical lattice.  A simple scaling relation of the quasi-two-dimensional loss rate constants with dipole strength, trap frequency and collision energy is given for like bosons or like fermions.   It should be possible to stabilize ultracold dimers of any of these species against destructive collisions by confining them in a lattice and orienting them with electric field of less than 20 kV/cm. }
\vspace{0.5cm}
 \end{@twocolumnfalse}
  ]

\section{Introduction}

\footnotetext{\textit{$^{a}$~Joint Quantum Institute , NIST and the University of Maryland, Gaithersburg, Maryland 20899-8423 USA Tel: +1-301-975-2596; E-mail: psj@umd.edu}}
\footnotetext{\textit{$^{b}$~Joint Quantum Institute , NIST and the University of Maryland, Gaithersburg, Maryland 20899-8423 USA. }}
\footnotetext{\textit{$^{c}$~Faculty of Physics, University of Warsaw, Ho{\.z}a 69, 00-681 Warsaw, Poland.. }}

Quite spectacular success has been achieved in recent years in working with gases or lattices of ultracold atoms cooled to temperatures on the order of 1 $\mu$K or less.  This has permitted the achievement of quantum degeneracy with either bosonic or fermionic isotopes of various atomic species where the thermal de Broglie wavelength becomes of the same order as or larger than the mean distance between atoms.  Quite precise control of the various properties of such systems is possible through state selection, trap design, and magnetic tuning of scattering resonances.  A number of reviews or books covering this work have appeared, including topics such as Bose-Einstein condensation,~\cite{Dalfovo1999, Stringari2003} the quantum properties of fermionic gases,~\cite{Giorgini2008,Pethick2008} and magnetically tunable Feshbach resonance control of collision properties.~\cite{Timmermans1999,Hutson2006,Hutson2007,Kohler2006,Chin2010}  In addition, much work has been directed towards creating lattice structures of such cold atoms using standing wave light patterns to make optical lattices of various geometric configurations.~\cite{Jessen1996,Bloch2005,Bloch2007,Greiner2008}   

Recent work has now succeeded in making ultracold molecules in their stable vibrational and rotational ground electronic state with temperature $T < 1$ $\mu$K by using magnetic and electromagnetic field control of ensembles of ultracold atoms.  This has been achieved for $^{40}$K$^{87}$Rb fermions\cite{Ni2008} and $^{133}$Cs$_2$  bosons~\cite{Danzl2010}, and for $^{87}$Rb$_2$ bosons in the collisionally unstable $v=0$ level of the lowest $^3\Sigma_u^+$ state.~\cite{Lang2008} These successful experiments used a two-step process to make the molecules.  First, magneto-association of two very cold atoms makes a very weakly bound ``Feshbach molecule'',~\cite{Kohler2006} which is then converted by a coherent STIRAP process  to make a ground state molecule in its $v=0$, $J=0$ ground state in a single state of nuclear spin, where $v$ and $J$ are the respective quantum numbers for vibration and rotation.  This builds on pioneering earlier work in much more dilute cold atomic gases around 100 $\mu$K to make ground state RbCs~\cite{Sage2005}, Cs$_2$,~\cite{Viteau2008} or LiCs~\cite{Deiglmayr2008a} molecules.   Cold molecules open up many new opportunities for study,~\cite{Demille2002,Doyle2004,Carr2009} since they have more complex internal structure and different long range potentials.  If dipolar, their properties and collisions can be controlled by an electric field in addition to magnetic or electro-magnetic fields.~\cite{Micheli2006,Buchler2007}

Having ultracold molecules also introduces the possibility of chemistry and reactions at ultralow energies, with precise control of the initial internal states and translational energy of the reactants.~\cite{Krems2008,Carr2009}  Having such collisions is good if one wishes to study such chemistry following the precise preparation of the initial states.~\cite{Julienne2009}  Collisions are bad, however, if one wishes to keep the molecules for simulating condensed matter systems or doing complex control like quantum computing, since reactive collisions can rapidly remove molecules from the gas in question.  Dipolar molecules offer some special features such as the possibility of orienting them by an electric field.  Reaction rates can be strongly suppressed if the molecules are oriented to have repulsive dipolar interactions while confined to move in a two-dimensional (2D) plane by an optical lattice wave guide.  Such suppression of reaction rates in this quasi-2D geometry has been both predicted~\cite{Micheli2007,Micheli2010,Quemener2010b} and demonstrated for $^{40}$K$^{87}$Rb.~\cite{Miranda2010}  Controlling internal spin can also be used to decrease reaction rates in the case of fermionic molecules like $^{40}$K$^{87}$Rb, since identical  fermions can only collide via odd partial waves that have centrifugal barriers to reaction.~\cite{Ospelkaus2010,Quemener2010,Idziaszek2010b}

Here we examine some basic aspects of ultracold chemistry of highly reactive molecules.  We restrict our considerations to the special case that the two interacting molecules have a unit probability of a chemical reaction or an inelastic quenching collision if they approach one another within typical chemical interaction distances, on the order of 1 nm or less.  Thus, we are considering the quantum threshold limit to the standard Langevin model.~\cite{Quemener2010,Idziaszek2010}  In this highly reactive limit  the long range potential between the molecules controls how they get together subject to experimental control on an ultralow energy scale.   This paper will examine various universal aspects of reaction rates that are solely governed by such long range interactions of two reactive molecules.~\cite{Idziaszek2010,Micheli2010}  Such interactions are sensitive to the Bose/Fermi character of the molecules, to the traps used to confine them, and to electric, magnetic, and electromagnetic fields in the case of polar molecules.  We will consider, in particular, the ten different molecular dimers comprised of two alkali metal atoms from the group Li, Na, K, Rb, and Cs.

The theory will be confined to the lowest temperatures where only the lowest partial waves allowed by symmetry contribute.  We will apply both analytic and numerical approaches to molecules colliding in the 3D geometry of free space or in the quasi-2D geometry of a confining optical lattice.  The basic theory related to ultracold molecules, their formation, states, collisions, dipolar properties, and response to external fields is discussed in detail in the introductory book by Krems {\it et al.}~\cite{Krems2009} and the Faraday Discussions 142.~\cite{Faraday2009}  We will review this theory as needed here and apply it to dipolar mixed alkali dimer molecules.

\section{Collisions in free space}

\subsection{Ultracold polar molecules}
\label{PolarMol}

While general cooling methods such as buffer gas cooling and Stark deceleration to load a molecule trap are being developed for a variety of molecules,~\cite{Doyle2004,Carr2009,Krems2009} such schemes so far have been restricted to temperatures significantly above 1 mK and very low phase space density, many orders of magnitude removed from quantum degeneracy.    To date, high phase space density has been restricted to molecules that can be made directly by associating two atoms that are at or near the quantum degenerate regime of temperature and density.  While initial proposals to do this involved photoassociation of the atoms,~\cite{Damski2003,Jaksch2002,Jones2006} it has turned out that magnetoassociation~\cite{Hutson2006,Kohler2006} provides an efficient and effective tool to make near threshold bound states known as Feshbach molecules,~\cite{Kohler2006,Chin2010} with binding energies $E/h$ less than 1 MHz.  These can  then be optically converted via a coherent Raman process to much more deeply bound vibrational levels,~\cite{Ospelkaus2008,Danzl2009} including the ground state.~\cite{Ni2008,Danzl2010}   Typical temperatures are well below 1 $\mu$K and densities can be on the order of $10^{12}$ molecules/cm$^3$ or even larger.  Since this can be done by coherent quantum dynamics that causes no heating, the molecules have the same temperature as the initial atoms.  It is even possible to make an optical lattice array of many single trapping cells,~\cite{Danzl2009,Danzl2010} in which exactly two atoms are present.  Upon associating the atoms, one then has an array of molecules, each of which is trapped in its own lattice cell, with confinement possible to tens of nm and intercell spacing of hundreds of nm.
 
In principle any diatomic molecule could be made that is comprised of atoms that can be cooled and trapped, including species such as Li, Na, K, Rb, Cs, Ca, Sr, Yb, or Cr.  In practice so far, ultracold molecule formation has been concentrated on alkali-metal atom dimers.  Among the alkali-metal-species Li, Na, K, Rb, and Cs, it is known~\cite{Zuchowski2010} that five mixed dimers have exoergic reactive collision channels when in their vibrational and rotational ground state, $v=0$, $J=0$, namely, LiNa, LiK, LiRb, LiCs, and KRb.  The other five, NaK, NaRb, NaCs, KCs, and RbCs, have no reactive channels for $v=0$, $J=0$.  

A molecule in a pure rotational eigenstate has no dipole moment.   Calculating the energy of a molecular dipole in an electric field $\bf{F}$ is explained in Chapter 2 of the book by Krems {\it et al.}~\cite{Krems2009}.  We apply this method to the case of a $^1\Sigma^+$ symmetric top rotor in vibrational level $v$ in its ground rotational state $J=M=0$, where $M$ is the projection of total angular momentum $J$ along the direction of $\bf{F}$.  Upon expanding the wave function in symmetric top basis states $|JM\Lambda v\rangle = |J00v\rangle$, where the body-frame projection $\Lambda=0$, the energy $E_g(F)$ of the lowest energy ground state $|g(F)\rangle$ of the manifold is found by diagonalizing the Hamiltonian matrix $H^\mathrm{mol}$ with matrix elements:   
\begin{eqnarray}
\label{Hfield1}
H_{JJ}^\mathrm{mol}&=&B_v J(J+1)\,\\
\label{Hfield2}
 H_{J,J+1}^\mathrm{mol} = H_{J+1,J}^\mathrm{mol}&=&\frac{J+1}{\sqrt{(2J+1)(2J+3)}} \, Ed_m \,,
\end{eqnarray}
where $B_v=\hbar^2\langle 000v|1/r^2|000v\rangle /M$ is the rotational constant for the molecule with mass $M$, $r$ is the interatomic separation, and $d_m$ is the body-frame permanent molecular dipole moment.    The energy $E_g(F)$ approaches the energy of the $v,J=0$ molecular level as $F \to 0$. The field-dependent dipole moment for the field-dressed ground state  is $d(F) = -\partial E_g(F)/\partial F$, which approaches $d(F) \to d_m \, (Fd_m)/(3B_v) $ as $F \to 0$.

\begin{figure}[h]
\centering
  \includegraphics[height=6cm]{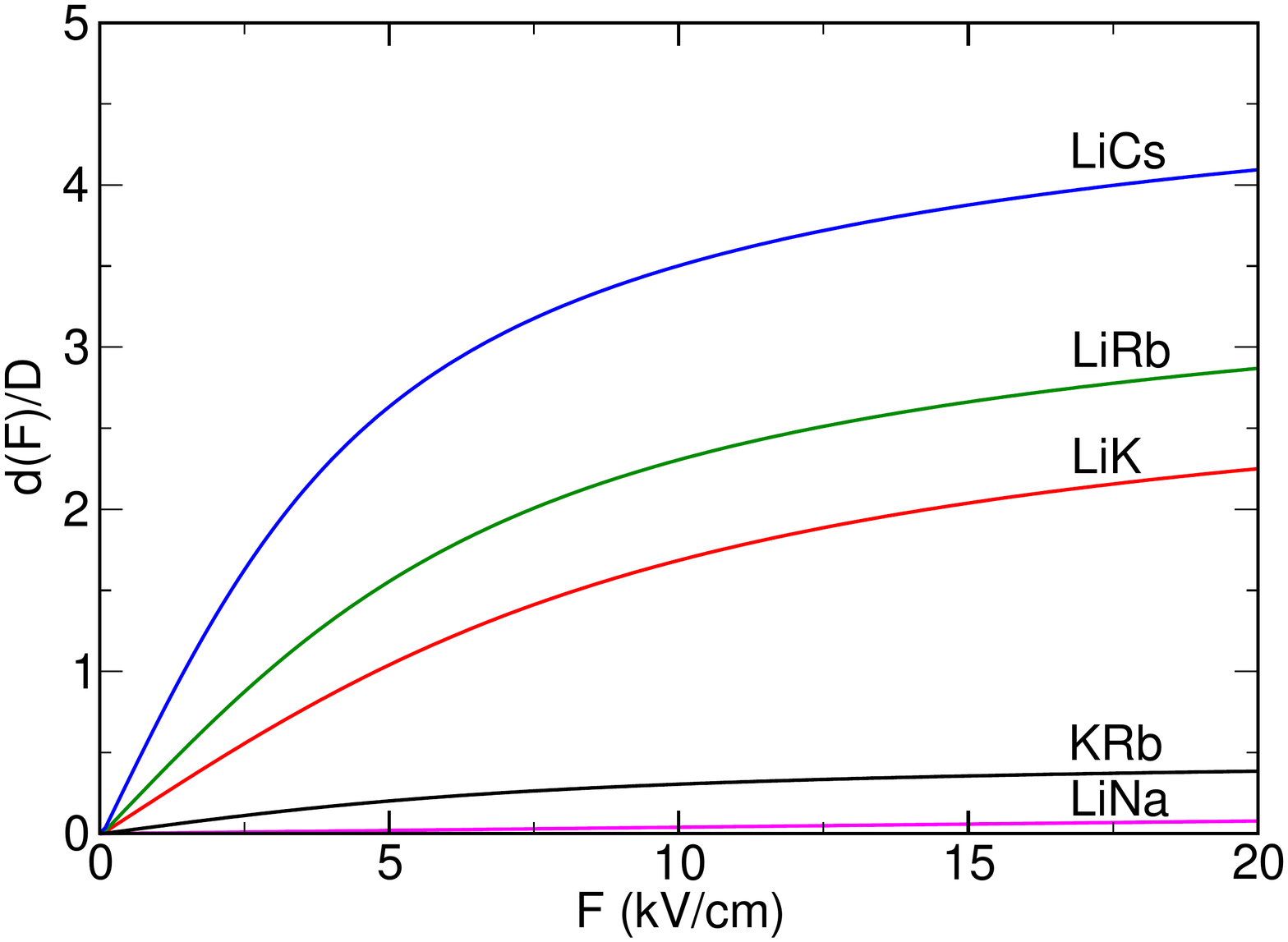}
    \includegraphics[height=6cm]{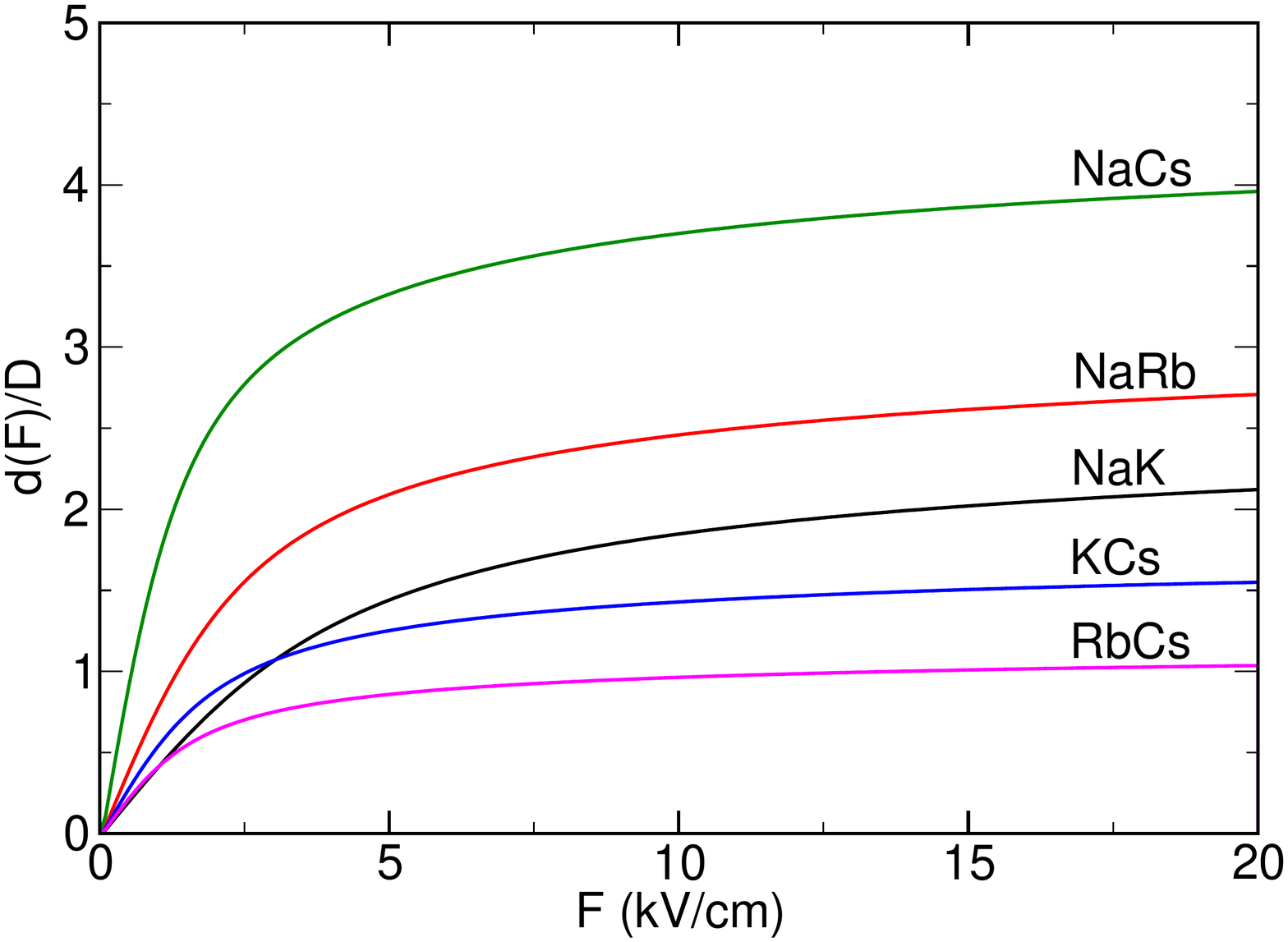}
  \caption{Dipole moment $d(F)/D$ versus $F$ for the five reactive mixed alkali-metal-species (upper panel) and for the five non-reactive mixed alkali-metal-species (lower panel), where $D = 0.3934$ au $=$ $3.336\times 10^{-30}$ Cm.}
  \label{fig1}
\end{figure}

In order to make estimates for the dipolar collision properties of the ground and lowest vibrational levels of the ten mixed-alkali-dimer species, we will take the dipole moments $d_m=d_e$ and rotational constants $B_e$ for the ground $^1\Sigma^+$ state evaluated at the equilibrium internuclear distance $R_e$  from the calculations of Aymar and Dulieu.~\cite{Aymar2005}  Figure~\ref{fig1}  shows the field-dependent dipole moments $d(F)$ as a function of $F$ for the respective reactive and non-reactive sets of species. 

A key aspect to note from Fig.~\ref{fig1} is that for all species except LiNa and KRb, dipole moments on the order of 0.4 au (1 D $=$ $3.336\times 10^{-30}$ Coulomb meter) can be achieved at relatively modest electric fields of 10 kV/cm or less.  As $F$ increases, $d(F)$ approaches and eventually reaches the magnitude of the molecular dipole $d_m$.  This is not yet achieved at the 20 kV/cm maximum in Fig.~\ref{fig1}.  The KRb experiments on collisions in an electric field were carried out with $F$ up to 5 kV/cm and were able to reach dipole moments only on the order of 0.08 au (0.2 D).~\cite{Ni2010,Miranda2010}  Except for LiNa, which has an even smaller dipole moment than KRb, all of the other species should be capable of being used with dipole moments on the order of 0.4 au (1D) or more to enable quite interesting experiments to control threshold collision dynamics.

\subsection{Characteristic energy and length scales}
\label{length}

In order to get a good understanding of the properties of ultracold molecule interactions, it is important to understand the various length scales $L$ and corresponding energy scales $\hbar^2/(2\mu L^2)$ associated with ultracold phenomena.  These are the de Broglie wavelength $2\pi/k$ and the lengths $R_e$, $\bar{a}$, $a_h$, and $a_d$ associated with the respective chemical, van der Waals, harmonic, and dipolar interactions.  For a relative collision kinetic energy of $E=\hbar^2 k^2/(2\mu)$, where $\mu=M/2$ is the reduced mass of the molecule pair of mass $M$, we define a characteristic energy-dependent length $a_k = 1/k$.  The characteristic van der Waals length is $\bar{a} = [2\pi/\Gamma(1/4)^2]\,(2\mu C_6/\hbar^2)^{1/4}$, where $-C_6/R^6$ is the van der Waals dispersion potential, $R$ is the intermolecular separation, and $2\pi/\Gamma(1/4)^2\approx 0.47799$.~\cite{Jones2006,Chin2010} The harmonic length for trapping frequency $\Omega$ is the characteristic length of ground state motion $a_h = \sqrt{\hbar/(\mu \Omega)}$.  The potential energy of interaction between two molecular dipoles separated by distance $R$ is $d(F)^2(1-3\cos^2\theta)/R^3$, where $\theta$ is the angle between dipole orientation and the intermolecular axis.  The dipole length $a_d(F) = \mu d(F)^2/\hbar^2$ is defined to be the length where $\hbar^2/(\mu a_d^2) = d(F)^2/a_d^3$.

In order to estimate $\bar{a}$, we need the $C_6$ value for the various species.     For a polar molecule there are two contributions to the effective $C_6$ interaction.  One is the electronic contribution $C_6^\mathrm{el}$, due to the second-order response through excited electronic states.  A simple Unsold approximation gives $C_6^\mathrm{el}=(3/4)U\alpha^2$, where $U\approx0.055(7)$ atomic units is a mean excitation energy and $\alpha$ is the dipole polarizability.~\cite{Azizi2004,Deiglmayr2008b} This gives a magnitude on the order of $C_6^\mathrm{el}\approx 10^4$ au (1 au $= E_h a_0^6=9.573\times 10^{-26}$ J nm$^6$)  for the mixed alkali dimers.  For most species, the much larger and dominant contribution to $C_6$ is the rotational dipole part with a magnitude $C_6^\mathrm{rot}=d_m^4/(6B_v)$.~\cite{Micheli2006}  Thus, for estimation purposes for the vibrational ground state, we use $C_6 \approx (3/4)U\alpha^2 + d_e^4/(6B_e)$.   This approximation can be compared to the calculations of  Kotochigova~\cite{Kotochigova2010} for KRb and RbCs, where our estimates give respectively $10500+2800$ au  and $15000+113000$ au for the two contributing terms, giving $C_6$ values 10 and 20 per cent less than the corresponding {\it ab initio} calculations.  Since the length $\bar{a}$ is not very sensitive to $C_6$, varying only as $C_6^{1/4}$,  this approximation makes small errors on the order of 5 per cent or less in our qualitative estimates of $\bar{a}$.  Consequently our estimates for universal van der Waals rate constants are likely to be in error by 15 per cent or less for all species except perhaps LiNa, which has the smallest dipole moment.

\begin{figure}[h]
\centering
  \includegraphics[height=6cm]{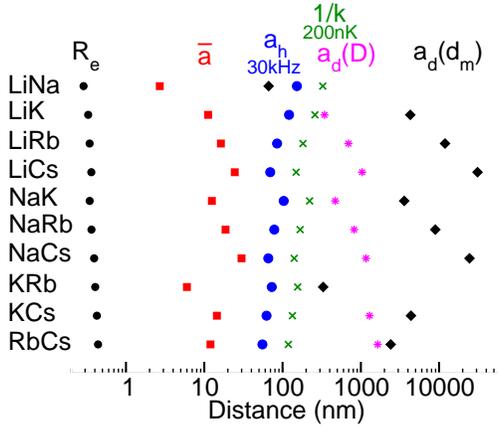}
  \caption{Characteristic length scales $R_e$ (black dots), $\bar{a}$ (squares), $a_h$ at $\Omega=2\pi(30\,\mathrm{kHz})$ (blue dots), thermal expectation value of $1/k$ at 200 nK (crosses), and $a_d$ (stars for $d=0.4$ au (1 D) and diamonds for $d=d_m$) for the ten mixed alkali-metal species.}
  \label{fig2}
\end{figure}

Figure~\ref{fig2} illustrates the various characteristic lengths for the ten mixed alkali dimers.  The shortest distance is the chemical bond length $R_e$, which is much less than 1 nm,~\cite{Aymar2005} with a corresponding energy scale $E/h$ on the order of the chemical bond energy of 100 THz.   The next largest energy scale, on the order of MHz, is that of the van der Waals length $\bar{a}$, which ranges between 6 nm for KRb and 30 nm for LiCs.  The harmonic confinement by an optical lattice typically has $\Omega/(2\pi)$ on the order of tens of kHz.  The Figure shows that the confinement $a_h$ is on the order of 100 nm for $\Omega = 2\pi(30\,\mathrm{kHz})$.  Figure~\ref{fig2} also shows the thermal average of $1/k$ is on the order of  a few hundred nm and is larger than either $\bar{a}$ or $a_h$ when collision energy $T$ is 200 nK, or $k_B T/h=4$ kHz, a typical value for ultracold systems.  Note that the product $k a_h=(2E/(\hbar \Omega))^{1/2}$ is independent of the mass of the species for a given trap frequency $\Omega$.  The trap-induced confinement $a_h$ is less than $1/k$ when $E < \frac{1}{2}\hbar \Omega$, or $E/k_B <$ 700 nK, for an $\Omega = 2\pi(30\,\mathrm{kHz})$ trap.

The dipole length depends on the electric field $F$.  Figure~\ref{fig2} shows $a_d(F)$ for two values of the dipole strength $d(F)$, namely, for 0.4 au (1 D) if this magnitude is possible, and for its maximum allowed value $d_m$.  These dipole length scales, respectively on the order of 1000 nm and 10000 nm, are the largest length scales in the problem, with the exception of LiNa, which has the smallest dipole moment.  At $d(F)=0.4$ au, the corresponding energy scales are on the order of 100 Hz, but over 1 kHz for LiNa, LiK and NaK.

\subsection{Universal collision rate constants}

The basic theory of ultracold collisions is covered in Chapters 1 and 3 of Krems {\it et al.}~\cite{Krems2009} and by Hutson~\cite{Hutson2007,Hutson2007b} and Chin {\it et al.}~\cite{Chin2010}.  We will summarize here the essential points using the formulation with quantum defect theory (QDT) by Idziaszek and Julienne~\cite{Idziaszek2010,Idziaszek2010b}  For the moment we will ignore the role of internal spin structure in the ground molecular state,~\cite{Aldegunde2008,Julienne2009} since we are considering highly reactive collisions where this seems not to be relevant.~\cite{Ospelkaus2010} The essence of the QDT approach is the separation of energy and length scales, as seen in Fig.~\ref{fig2}.  The overall effect of the short range ``chemistry'' zone with $R \ll \bar{a}$ is summarized in the outer van der Waals zone by two dimensionless QDT parameters $s$ and $0 \le y \le 1$, which respectively represent a short range phase and the chemical reactivity in the entrance channel $j$; the scattering length of the entrance channel  is parameterized by $s$, which can take on any value, and $y=1$ represents unit probability of short range loss from the entrance channel, whereas $y=0$ means no loss.  In the ultracold domain, we only need to consider the lowest partial wave $j$ to index the channel, where $j=0$ for $s$-waves for like bosons or for unlike bosons or fermions, and $j=1$ for $p$-waves for like fermions or for unlike bosons or fermions.    

The case of $y=1$ corresponds to a special ``universal'' case, where there is unit probability of short range loss.  In this case, the elastic and inelastic or reactive collision rates  depend only on the quantum scattering by the long range potential at distances $R \gtrsim \bar{a}$.  Any incoming scattering flux that penetrates inside $\bar{a}$ experiences no reflection back into the entrance channel since it is lost to reactive or inelastic channels.  No scattering resonances can exist in this case.  The rate constant for loss from the entrance channel $K_j^\mathrm{ls}(E)$ for a relative collision kinetic energy $E$ for a van der Waals potential takes on the following very simple universal form, independent of the details of the short range potential or dynamics, at low collision energy where $k\bar{a} \ll 1$,
\begin{equation}
\label{}
   K_0^\mathrm{ls} = g\frac{4\pi \hbar}{\mu} \bar{a} \,\,\,\mathrm{and}\,\, \,K_1^\mathrm{ls}(E) =3g \frac{4\pi \hbar}{\mu}\,(k\bar{a})^2\bar{a}_1\,,
\end{equation}
where the identical particle factor $g=2$ if the collision partners are identical bosons or identical fermions, and $\bar{a}_1=\bar{a}\Gamma(\frac{1}{4})^6/(144\pi^2\Gamma(\frac{3}{4})^2)\approx 1.064 \bar{a}$.~\cite{Idziaszek2010}   If the colliding particles are not identical, then the rate constant is $K^\mathrm{ls}=K_0^\mathrm{ls} + K_1^\mathrm{ls}(E)$ with $g=1$, and the second term becomes much smaller than the first as $k \to 0$. According to the thresold law for a van der Waals potential, the thermal average is independent of temperature $T$ for the $s$-wave but varies linearly with $T$ for the $p$-wave,
\begin{equation}
\label{ }
K_1^\mathrm{ls}(T)= 1513\,\bar{a}^3\,k_BT/h \,,
\end{equation} 
where the factor for identical fermions comes from $\Gamma(\frac{1}{4})^6/\Gamma(\frac{3}{4})^2\approx 1513$.  The density $n$ of a uniform gas of identical bosons or fermions varies as $\dot{n}=- K_j^\mathrm{ls}(T) n^2$.  In the case of a two-species gas, the two densities $n_1$ and $n_2$ vary as $\dot{n}_1=\dot{n}_2=-K^\mathrm{ls} n_1 n_2$.

\begin{figure}[h]
\centering
  \includegraphics[height=6cm]{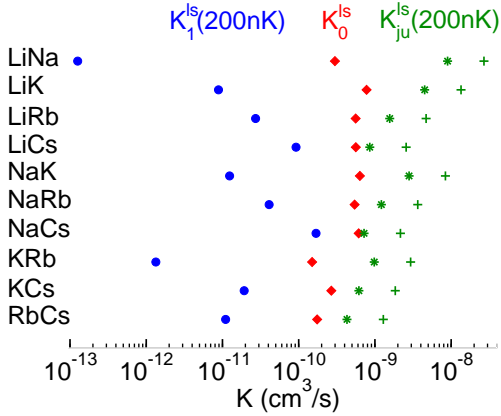}
  \caption{Universal rate constants for like bosons, $K_0^\mathrm{ls}$( red diamonds), like fermions $K_1^\mathrm{ls}$ (blue dots), and the unitarity limit, $K_{ju}^\mathrm{ls}$ (green star for $j=0$, green plus for $j=1$), for van der Waals collisions at $T=200$ nK for the ten mixed alkali-metal species.  These are expected to give the correct order of magnitude of $K_j^\mathrm{ls}$ for reaction of $v=0,J=0$ for the five reactive species LiNa, LiK, LiRb, LICs, KRb and for vibrational quenching of low $v \ge 1$ levels for the five species that are not reactive for $v=0$, NaK, NaRb, NaCs, KCs, and RbCs.  As $T \to 0$, $K_1^\mathrm{ls}(T)$ varies linearly with $T$, $K_{ju}^\mathrm{ls}(T)$ varies as $1/\sqrt{T}$, and $K_0^\mathrm{ls}$ is independent of $T$. As $T$ increases the universal rate limits only apply as long as $K_j^\mathrm{ls}(T) < K_{ju}^\mathrm{ls}(T)$.}
  \label{fig3}
\end{figure}

The universal rate will only apply to species with $y=1$ and thus to all five of the highly reactive species in the upper panel of Fig.~\ref{fig1}.  This has been demonstrated for KRb for $T < 1$ $\mu$K.~\cite{Ospelkaus2010}  However, there is good evidence that excited vibrational levels of alkali dimer molecules collisionally quench to lower vibrational levels with near-universal rate constants on the order of $10^{-10}$ cm$^3/$s.  This is suggested by theoretical calculations~\cite{Quemener2005,Quemener2007} on collisions of alkali-metal atoms with alkali-metal dimers, and by experimental measurements on Cs with Cs$_2$~\cite{Staanum2006,Deiglmayr2011}, Cs and Rb with RbCs~\cite{Hudson2008}, and Cs with LiCs.~\cite{Weidemuller2010}  The experiments are at higher $T$ where more than the lowest partial wave may contribute.  The RbCs work~\cite{Hudson2008} included a specific theoretical calculation of the $T$-dependent universal rate constants summed over partial waves that agreed with the measured quenching rate constants.  Therefore, we will calculate the universal rate constants for the five non-reactive species in the lower panel of Fig.~\ref{fig1} and assume that they give the order of magnitude of the vibrational quenching rate constant in the $T \to 0$ limit for alkali-metal dimer molecules in states with $v \ge 1$ due to a collision with an alkali-metal atom or another dimer molecule.  Future experiments can check whether this assumption gives good approximate magnitudes for the actual quenching rate constants.

Figure~\ref{fig3} shows the universal rate constants $K_0^\mathrm{ls}$ and $K_1^\mathrm{ls}(T=200\mathrm{nK})$ for identical bosons and fermions respectively, as well as the $s$-wave unitarity limit
\begin{equation}
K_{0u}(T)=g(\pi\hbar/\mu)\langle1/k\rangle_T\,,
\end{equation}
where $\langle1/k\rangle_T$ represents a thermal average of $1/k$.  The unitary limit $K_{ju}(T)$ gives the upper bound on the rate constant.  Consequently the universal rate constants only apply if they are smaller than this limit, requiring $k\bar{a} \lesssim 1/4$ for $j=0$ and $k^3\bar{a}^3 \lesssim 1/4$ for $j=1$.  The corresponding $p$-wave unitarity limit $K_{1u}(T)$ for like fermions is 3 times larger than $K_{0u}(T)$ for like bosons.  

All five alkali-metal species have bosonic isotopes, but only Li and K have stable fermionic isotopes.  Consequently stable NaRb, NaCs, and RbCs fermions do not exist, but $^{22}$Na and $^{134}$Cs have half lives longer than 2 years.  The fermionic molecules tend to have rate constants much less than the unitarity limit at this $T$.  The bosonic molecules are closer to their unitarity limit for 200 nK, especially those with the highest dipole moments.

Since the decay rate per molecule for a gas of density $n$ is $K_j^\mathrm{ls} n$, the lifetime of the molecule is $\tau=(K_j^\mathrm{ls} n)^{-1}$.  Thus, taking $n=10^{12}$ cm$^{-3}$ as a typical ultracold gas density, one gets lifetimes of 1 s and 1 ms for respective rate constants of $10^{-12}$ cm$^3/$s and $10^{-9}$ cm$^3/$s.  Thus reactive bosons and unlike fermions will have typical lifetimes of a few ms and the less reactive identical fermions will have typical lifetimes in the 10 ms to 100 ms range for $T$ around 200 nK, depending on species.  Two exceptions are KRb and LiNa fermions, which have relatively small dipole moments.  A lifetime on the order of 1 s has been achieved for like KRb fermions in the 200nK range.~\cite{Ospelkaus2010}  The identical fermion lifetime will increase as $1/T$ as $T$ decreases.

\subsection{Effect of an electric field} 

When an electric field is turned on, the reaction rate constant for like fermions increases towards unitarity with increasing field strength.  This has been explained quite simply by Qu{\'e}mem{\'e}r and Bohn,~\cite{Quemener2010,Ospelkaus2010} who adapted a Langevin model to quantum threshold conditions.  This can be done more rigorously through a quantum defect model of the threshold barrier penetration probability.~\cite{Idziaszek2010b} Detailed analysis of free space collisions of two $J=0$ $^1\Sigma$ dipolar molecules have been treated quite well in these references and others.~\cite{Ticknor2005,Ticknor2007,Ticknor2008,Bohn2009}  
Here, we discuss approximations suitable for universal ultracold collisions of such dipolar molecules as the electric field $F$ is turned on. As collision energy increases above the threshold regime,  universal aspects of dipolar collisions also appear.~\cite{Bohn2009} Since more than one partial wave can contribute to the collision rate as $F$ increases, we include a sum over all partial waves.  This is important for ultracold elastic collisions, although ultracold reaction or loss collision rates tend to be dominated by the contribution from the lowest partial wave in our range of electric field strengths.

Since scattering flux is incoming only inside the van der Waals radius $\bar{a}$ for such collisions and there are no scattering resonances, the details of the anisotropic potential for $R \ll \bar{a}$ are not important.  The collision rates are determined by quantum scattering by the anisotropic long-range potential, which needs to be fully treated in a numerical approach that couples different $|JM0v\rangle$ rotational states through the electric field and the anisotropic potential.  For this purpose we use a coupled channels expansion in a $|JM0v\rangle$ basis.  For the present we consider only collisions of the lowest Stark level that correlates with the $J=M=0$ molecular state for $F=0$.   When $R$ is very large, the Stark shift in energy levels is very small compared to the spacing $2B_v J$ to the $J=1$ rotational level.  The dominant contributions to the potential are the direct dipole-dipole interaction, varying as $R^{-3}$, and an isotropic second-order dispersion interaction  given by a van der Waals constant $C_6(F)$ that depends on $F$.  When $R$ becomes small enough, at distances $R \ll \bar{a}$, the strong dipole-dipole interaction becomes larger than $2 B_v J$, and a perturbation treatment of the dispersion interaction breaks down.  Effectively, the two interacting dipoles become more strongly coupled to one another than to the imposed electric field $F$.  A coupled channels expansion is neded to account for this changing coupling during the course of the collision.

The interaction part of the Hamiltonian for the coupled channels expansion reads
\begin{equation}
\label{Hrot}
H_\mathrm{int} = \sum_{j=1}^{2} H_{j}^\mathrm{mol} + V_\mathrm{dd}(\mathbf{R}) + V_\mathrm{el}(\mathbf{R}),
\end{equation}
where $H_j^\mathrm{mol} = B_\nu \mathbf{J}_j^2 + \mathbf{F} \cdot \mathbf{d}_j$ is the Hamiltonian of a single  molecule defined in Sec.~\ref{PolarMol}, with $\mathbf{J}_j$ and $\mathbf{d}_j$ denoting the angular momentum and the dipole moment operators of molecule $j=1,2$ respectively. The dipole-dipole interaction
\begin{equation}
V_\mathrm{dd}(\mathbf{R}) = \frac{\mathbf{d}_1 \cdot \mathbf{d}_2 - 3 (\mathbf{d}_1 \cdot \mathbf{e}_R) (\mathbf{e}_R \cdot \mathbf{d}_2)}{R^3}
\end{equation}
depends on the projection of the dipole operator $\mathbf{d}_j$ on the axis $\mathbf{e}_R = \mathbf{R}/R$ connecting the center of masses of the two molecules.
Finally, $V_\mathrm{el}(\mathbf{R})$ describes the electronic contribution to the dispersion interaction between molecules, given in Section~\ref{length} as $V_\mathrm{el}(\mathbf{R}) = - C_6^\mathrm{el} R^{-6}$ with $C_6^\mathrm{el}$ given by an Unsold approximation.

The most complete approach is to expand the full Hamiltonian, including the kinetic energy operator, in the basis $|J_1 M_1 0 v\rangle |J_2 M_2 0 v\rangle |\ell m \rangle$ of eigenstates of the symmetric top $|J_j M_j 0 v \rangle$ associated with a molecule $j$ in level $v$ and the eigenstates of the  angular momentum $\mathbf{L}$ of the relative motion $|\ell m \rangle$. Here, $\ell$ is the partial wave quantum number and $m$ denotes the projection of  $\mathbf{L}$ on the quantization axis in the laboratory frame.  This Hamiltonian conserves the projection of the total angular momentum on the electric field axis $M_\mathrm{TOT} = M_1 + M_2 + m$, and in the absence of the electric field also the total angular momentum $\mathbf{J}_\mathrm{TOT} = \mathbf{J}_1 + \mathbf{J}_2 + \mathbf{L}$. The matrix elements of the $H_{j}$ and the $V_\mathrm{dd}$ terms have been extensively discussed in the literature,~\cite{Micheli2007,Bohn2009b} and we do not need to give explicit formulas here.  

Typically the number of channels that have to be included in the quantum dynamics in the full rotational basis is very large. For example, the number of eigenstates with angular momentum $J_i \leq J_\mathrm{max} = 5$ and partial wave quantum numbers $\ell \leq \ell_\mathrm{max} = 10$ with $M_\mathrm{TOT}=0$ and a symmetry of the wave function corresponding to bosons is approximately 5700. This makes the coupled channels calculations highly computationally demanding. Hence, in our study of the effects of the full rotational expansion we apply the adiabatic approximation, diagonalizing the interaction part of the Hamiltonian at different values of the intermolecular distance $r$ and electric field $F$. This is motivated by the fact that the collisions of polar molecules in the universal $y=1$ regime exhibit no resonances, and in free-space can be accurately modeled in the framework of the adiabatic potentials.~\cite{Idziaszek2010b}  Our calculations with $n$ coupled adiabatic channels are designated $n$-channel adiabatic (rotational basis), and here we consider only $n=1$.

\begin{figure}[h]
\centering
 \includegraphics[height=5cm]{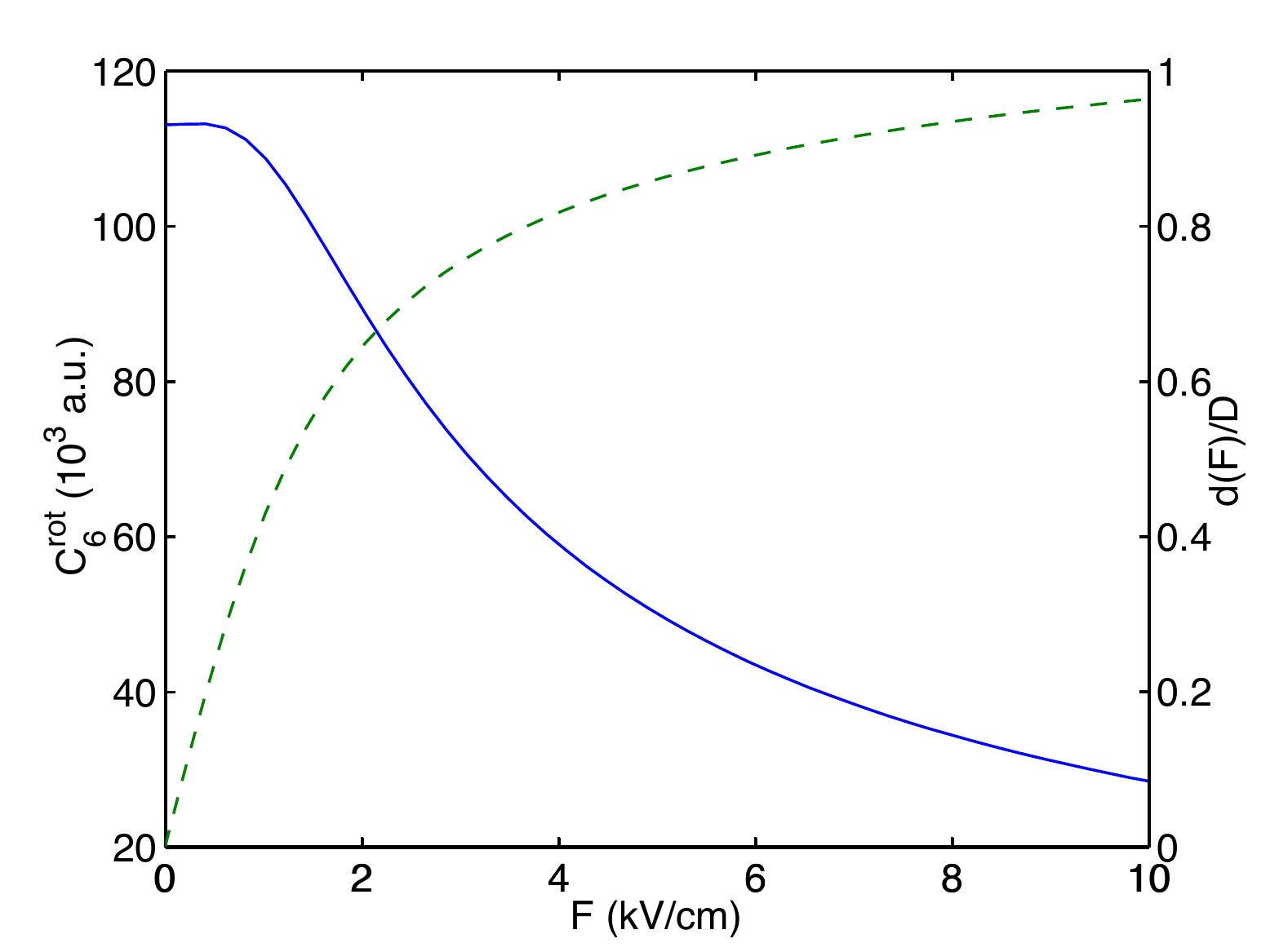}
 \caption{Field-dependent rotational contribution $C_6^\mathrm{rot}(F)$ to the van der Waals coefficient for the interaction of two ground state RbCs molecules (solid blue line) as a function of electric field $F$. Also shown is the dipole moment (dashed green line). }
 \label{fig4}
\end{figure}

When the molecules are far apart,  they interact very weakly with one another and each molecule is described by a field-dressed state $|g(F)\rangle$ with energy $E_g(F)$ and dipole $d(F)$ along the field direction, found by diagonalizing the single molecule Hamiltonian described in Section~\ref{PolarMol}.   At long range $H_\mathrm{int}$ in Eq.~(\ref{Hrot}) can be replaced by
\begin{equation}
\label{Vdisp}
V_\mathrm{int}(R,\theta) = - \frac{C_6(F)}{R^6} + d(F)^2 \frac{1 - 3 \cos^2 \theta}{R^3}
\end{equation}
where $\theta$ is the angle between the axis of the electric field and the intermolecular axis $\mathbf{R}$, and we take the zero of energy to be $E_g(F)=0$ for a given field $F$.  The dispersion coefficient $C_6(F) = C_6^\mathrm{el} + C_6^\mathrm{rot}(F)$  includes a field-dependent rotational contribution which can be evaluated from the second-order degenerate perturbation theory formula (see, e.g., Ref.~\cite{Kotochigova2010}):
\begin{equation}
- \frac{C_6^\mathrm{rot}(F)}{R^6}  = -\sum_{e \ell' m_\ell'} \frac{\langle g \ell m_\ell | V_\mathrm{dd} | e \ell' m_\ell' \rangle 
\langle e \ell' m_\ell' | V_\mathrm{dd} | g \ell m_\ell \rangle }{E_e - E_g} \, .
\label{eq:c6}
\end{equation}
Here, $g$ and $e$  respectively represent field-dependent ground and excited product states of the two molecules from the spectrum of solutions to $H_j^\mathrm{mol}$.  Symmetrizing appropriately, we decompose the single-molecule field dressed states into a $|JM0v\rangle$ basis, and then calculate the matrix elements of $V_\mathrm{dd}$ using the approach of Chapter 2 of Ref.~\cite{Krems2009}  We evaluate the sum over rotational states $J$ and partial waves $\ell$, and sum over projections such that $M_\mathrm{TOT}$ is conserved. We neglect the other degrees of freedom, which contribute to the non-rotational part of $C_6$, and have been calculated for some species through full electronic structure calculations.~\cite{Kotochigova2010}  Figure~\ref{fig4} gives an example of $C_6^\mathrm{rot}(F)$ for the $v=0,J=0$ ground state of $^{87}$Rb$^{133}$Cs.  The Figure shows that $C_6^\mathrm{rot}(F)$ decreases appreciably with increasing $F$, due mostly to increasing size of the energy denominators in Eq.~(\ref{eq:c6}).

A description in terms of the dispersion potential in \eqref{eq:c6} breaks down at small distances, when the dipole-dipole interaction becomes larger than $2 B_\nu J$. The characteristic length at which this occurs is $a_B = \left(d_m^2/B_\nu\right)^{1/3}$.~\cite{Micheli2007} This length is on the order of $0.4\bar{a}$ to  $0.6\bar{a}$ for the molecules we consider here.  Consequently, since $a_B$ is already effectively in the short range regime where all incoming flux of particles is lost, a description of collisions using Eq.~(\ref{Vdisp}) should be valid.  We have carried out coupled channels (CC) calculations to test this, by using a partial wave expansion in $|\ell m\rangle$ states with the full Hamiltonian including the interaction term in the form of Eq.~(\ref{Vdisp}).  We call this method the CC (vdW $+$ dipole) calculation.  In addition, we have implemented an adiabatic approximation in this expansion, designated either 1-channel adiabatic (vdw $+$ dipole) or 2-channel adiabatic (vdw $+$ dipole).

\begin{figure}[h]
\centering
 \includegraphics[height=5cm]{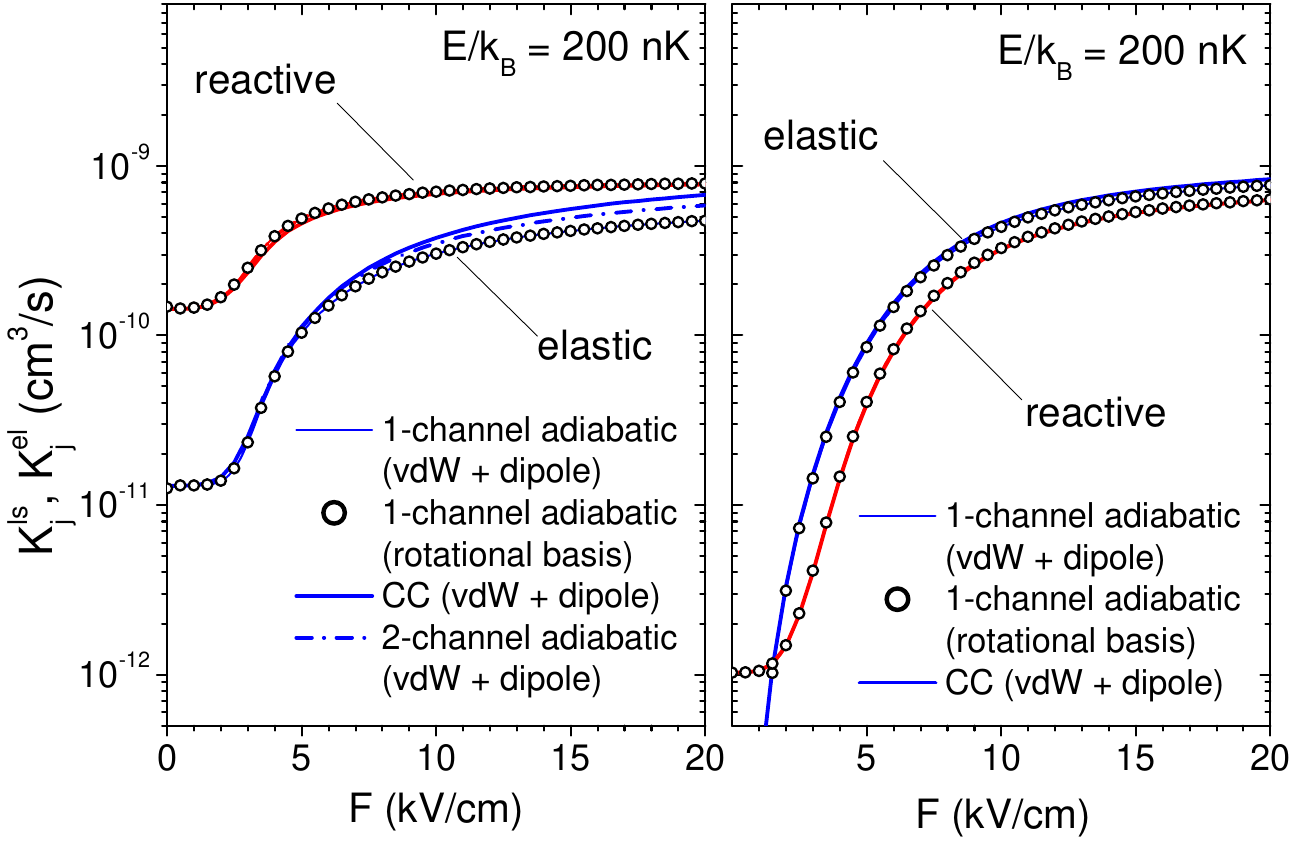}
 \caption{Field-dependent rate constants $K_0^\mathrm{ls}(F)$ and $K_1^\mathrm{ls}(F)$ at $E/k_B=200\mathrm{nK}$ versus electric field $F$ for free space collisions of KRb bosons (left panel) and fermions (right panel).   The figure shows the CC (vdW $+$ dipole) calculation (solid lines) and approximations based on the 1-channel adiabatic (rotational basis) (points), 1-channel adiabatic (vdw $+$ dipole) (thin solid line), and the 2-channel adiabatic (vdw $+$ dipole) approximations (dot-dash line). }
 \label{fig5}
\end{figure}

Figure~\ref{fig5} shows the elastic and reactive rates calculated with several approximations applied to free space collisions of bosonic and fermionic KRb prepared in the $F$-dependent state that correlates with the $v=0,J=0$ state as $F \to 0$.  For simplicity, we assume the same reduced mass and the dipole moment for bosons as for fermionic $^{40}$K$^{87}$Rb.  The approximations are: the CC (vdw $+$ dipole), the 1- and 2-channel adiabatic (vdw $+$ dipole) and the 1-channel adiabatic (rotational basis).  All approaches predict almost identical rate constants, except for the elastic rates for bosons at high electric fields. In this case the rates given by the CC (vdw $+$ dipole) are larger than those calculated using the adiabatic approximations. The discrepancy stems from the $\ell$-changing transitions induced by the off-diagonal elements in the dipole-dipole interaction, which are absent in the 1-channel adiabatic approximation. At high values of the electric field the elastic rates for bosons contain a significant contribution from the higher partial waves. This contribution can be included by considering more than one adiabatic potential. This is shown in Fig.~\ref{fig5} for bosons using the 2-channel adiabatic (vdw $+$ dipole) approximation, which goes much of the way to eliminating the difference between the adiabatic and coupled channels approaches.  Interestingly we have found numerically that calculated rates change by an insignificant amount if we use $C_6(0)$ instead of $C_6(F)$.  Presumably this is because the van der Waals dispersion term has become very small in comparison to the centrifugal and dipolar terms in the range that determines the dynamics.

\section{Collisions in quasi-2D geometry}

\subsection{Optical lattices}

The very low kinetic energy of ultracold atoms and molecules allows them to be controlled and manipulated by very weak forces.  One important control is provided by optical lattices, by which two counter-propagating light beams in the $z$ direction define a standing wave light pattern that defines a series of lattice cells separated by distances $L$ in the $z$ direction on the order of hundreds of nm to more than a $\mu$m.  We consider a single isolated cell with local harmonic confinement  along $z$ with frequency $\Omega/(2\pi)$ on the order of tens of kHz.  If $k_B T \ll \hbar \Omega$, the atoms or molecules are tightly confined to the ground state of the cell in the $z$ direction, while they are free to move in the $x-y$ plane.  Collisions under such conditions are called quasi-2D collisions.  On the other hand, two orthogonal sets of counterpropagating light beams can provide tight confinement in the $x$ and $y$ directions, with free motion in the $z$ direction.  This gives rise to quasi-1D collisions.  The theory of quasi-2D and quasi-1D collisions, and their threshold laws, have been worked out in some detail,~\cite{Olshanii1998,Sadeghpour2000,Petrov2001,Bergeman2003,Naidon2006,Naidon2007,Li2008,Li2009} including collisions of ultracold molecular dipoles in quasi-2D.\cite{Buchler2007,Micheli2007,Micheli2010,Quemener2010b,Quemener2011,DIncao2011} 

The big advantage of reduced dimensional collisions in an optical lattice is the extra control one gets over collision rates with dipolar molecules, since the dipole length $a_d$ can easily exceed the confinement length $a_h$.  Consequently, electric fields can produce oriented molecules and control their approach at long range, making it either attractive or repulsive.  In the latter case, short range reactive or loss collisions can be effectively turned off, thus greatly increasing the lifetime of the sample.

\subsection{Collision rates in reduced dimension}

Here we will give examples of the universal collision rate constants for quasi-2D collisions, when the molecules collide in an electric field ${\bf F}$ in the direction $z$ of tight confinement by a 1D optical lattice.  The species $^{40}$K$^{87}$Rb has already been treated in some detail for such a case,~\cite{Quemener2010b,Micheli2010,Quemener2011} and the predicted suppression of reactive collision rates by the repulsive dipole potential barrier has been verified experimentally.~\cite{Ni2010}  Here we examine the transition from van der Waals dominated to dipole-dominated collisions for the other species of mixed alkali-metal dimers.  The van der Waals limit applies when $a_d \ll \bar{a}$, whereas the dipolar limit applies when $a_d \gg a_h$.  In the latter case, collision rates approach a universal dipolar limit, as described in the literature.~\cite{Bohn2009,Ticknor2009,Ticknor2010,DIncao2011}

We use here the form of the elastic and inelastic collision rates from Micheli {\it et al.}~\cite{Micheli2010}, since this makes it easy to relate collisional properties in different dimensions $N$, where $N=1,2,3$ refers to quasi-1D, quasi-2D, and free space collisions respectively.~\cite{Naidon2006}    The ``core'' of the collision at short distances $ R \ll \bar{a}$ is assumed to occur in full 3D geometry and to be represented by a normal spherical harmonic expansion in partial waves.  The long-range part of the collision occurs in reduced dimensionality due to the tight confinment in 1 or 2 dimensions  with characteristic length $a_h \gg \bar{a}$ (see Fig.~\ref{fig3}). The kinetic energy $E=\hbar^2 \kappa^2/(2\mu)$ of free motion in the respective $(z)$, $(x,y)$, or $(x,y,z)$ directions for $N=1,2,3$ is assumed to be $ \ll \hbar \Omega$, so the colliding molecules are confined to the ground state of the harmonic trap.  This assumption can be easily relaxed to permit collisions of molecules in other trap levels.~\cite{Ni2010,Quemener2010}

The wave function can be expanded in a set of partial waves $j$ suitable for each dimension.~\cite{Naidon2006} The contributions to the elastic and inelastic rate constants for collisions with relative momentum $\hbar \kappa$ in dimension $N$ are~\cite{Micheli2010}
\begin{eqnarray}
  K_j^\mathrm{el}(\kappa)&=& \frac{\pi \hbar}{\mu} g_{N} \frac{|1-S_{jj}(\kappa)|^2}{\kappa^{N-2}} \,\,\,  \label{KNel} \\
   K_j^\mathrm{ls}(\kappa) &=& \frac{\pi \hbar}{\mu} g_{N} \frac{1-|S_{jj}(\kappa)|^2}{\kappa^{N-2}} \,, \label{KNls}
\end{eqnarray}
where the factor $g_N = 1/\pi\,,2/\pi\,,2$ for molecules colliding in identical spin states in $N=1,2,3$ dimensions, respectively.  The indices of the lowest partial wave are $j=0$ for bosons and $j=1$ for  fermions.   In 3D, $j=1$ has 3 components of its projection $M$ that have to be summed over.  In quasi-2D with $\bf{F}$ along the confined direction $z$, $j$ refers to the projection $M$ of relative rotational angular momentum along $z$, with two components $+1$ and $-1$ that have to be summed over for $j=1$; in quasi-1D, it refers to the symmetric ($j=0)$ or antisymmetric ($j=1$) state propagating along $z$.  In the case of unlike species both $j=0$ and 1 contribute.  The contribution from the lowest index is dominant for ultracold universal reactive collisions in an electric field, and any additional partial waves can be summed over in other cases to get the total rate constants $K^\mathrm{ls}$ or $K^\mathrm{el}$.  The loss rate for a gas with $N$-dimensional density n is $\dot{n}=-K^\mathrm{ls} n^2$, where $n$ has units of cm$^{-1}$, cm$^{-2}$, and cm$^{-3}$ for $N=1,2,3$ respectively.  

Micheli {\it et al}~\cite{Micheli2010} worked out analytic expressions for any $N$ for the universal $K^\mathrm{ls}$ or $K^\mathrm{el}$ for a van der Waals potential.  The loss rate constants for $N=2$ are $K_0^\mathrm{ls}=2(\hbar/\mu) P_0$ for $M=0$ bosons and $K_1^\mathrm{ls}=4(\hbar/\mu) P_1$ for fermions summed over both $|M|=1$ components.  The prefactor $2(\hbar/\mu)$ or $4(\hbar/\mu)$ gives the unitarity upper bound, and the transmission function $P_j=1-|S_{jj}|^2$ gives the dynamical probability of getting from the asymptotically prepared entrance channel to small $R$, where loss occurs:
\begin{equation}
\label{eq:KvdW}
P_0 = 4\sqrt{\pi} \frac{\bar{a}}{a_h}\,f_0(\kappa)  {\hspace{1 cm}} P_1 = 6\sqrt{\pi} \frac{\bar{a}_1}{a_h}\,(\kappa \bar{a})^2\,,
\end{equation}
where $f_0(\kappa)$ is a complicated function that has a logarithmic singularity as $\kappa \to 0$.  However, this singularity is not important in the experimental nK range (it is only dominant at much lower energies for typical trapping geometries).  For practical purposes $f_0(\kappa)$ can be taken to be on the order of unity and nearly independent of $\kappa$ for likely current experiments.  

We calculate $K^\mathrm{ls}$ here for quasi-2D geometry as a function of electric field $F$ using the numerical method described by Micheli {\it et al.},~\cite{Micheli2010} similar to the method of Qu{\'e}men{\'e}r and Bohn.~\cite{Quemener2010b,Quemener2011}  In our case of universal collisions, we impose incoming-only boundary conditions on the wave function for $R \ll \bar{a}$, which corresponds to loss of all scattering flux that reaches this region from long range.~\cite{Idziaszek2010,Micheli2010}  The calculation thus requires only the van der Waals, dipolar, trapping, and centrifugal potentials, with a spherical basis expanded in spherical harmonics at short range, switching to a cylindrical basis of confined states at long range.  The details of short range ``chemistry'' are irrelevant, since only loss from the entrance channel occurs there, due to reaction of vibrational quenching.  Since there is no back reflection to long range in the entrance channel, scattering resonances do not exist in this universal case.

\begin{figure}[h]
\centering
  \includegraphics[height=7cm]{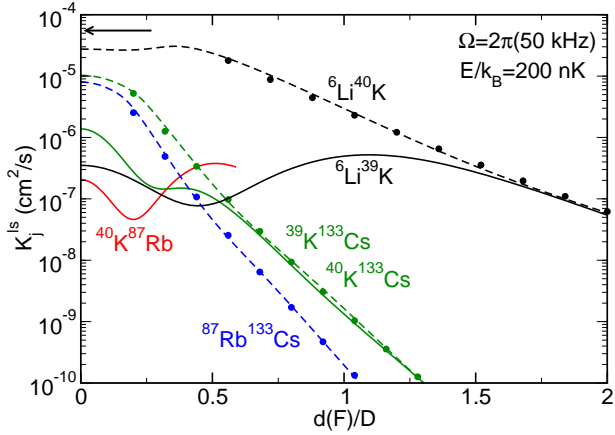}
  \caption{Universal loss rate constant $K_0^\mathrm{ls}$ for quasi-2D collisions at $E/k_B=200$ nK in an $\Omega=2\pi(50\,\mathrm{kHz})$ trap for bosonic ($j=0$) $^6$Li$^{40}$K,  $^{39}$K$^{133}$Cs and $^{87}$Rb$^{133}$Cs (dashed lines) and $K_1^\mathrm{ls}$ for fermionic  $^6$Li$^{39}$K, $^{40}$K$^{87}$Rb and $^{40}$K$^{133}$Cs (solid lines).    In the case of the nonreactive species KCs and RbCs, these rate constants are expected to apply to vibrational quenching for low vibrational levels with $v \ge 1$.  The black arrow shows the unitarity limit $2\hbar/\mu$ for bosonic  $^6$Li$^{40}$K.  The solid points show the fit function in Eq.~\ref{eq:KSC} when $a_d > a_h$.  }
  \label{fig6}
\end{figure}

Figure~\ref{fig6} shows our calculations of $K_j^\mathrm{ls}$ as the electric field $F$ is increased for a variety of species at a collision energy of $E/k_B=$ 200 nK for confinement in an $\Omega=2\pi(50\,\mathrm{kHz})$ trap.   These go to the universal van der Waals limits as $F \to 0$.  As we found for free space collisions, our numerical calculations change by an insignificant amount if we use $C_6(0)$ instead of $C_6(F)$ as $F$ increases.   Consequently the van der Waals potential becomes progressively less important as $F$ increases and the dipole potential becomes dominant on distance scales larger than the confinement length $a_h$.  We also find that for the low energies and moderate dipole strengths that we explore in quasi-2D geometry, the lowest partial wave is the dominant contribution to the universal loss rate constant, so little change would be seen in $K_j^\mathrm{ls}$ in Fig.~\ref{fig6} if the contributions from higher M values were included.  Elastic collisions, however, will be affected more, but they can be treated by simple approximations.\cite{Bohn2009,Ticknor2009}

The $^{40}$K$^{87}$Rb case in Fig.~\ref{fig6} is the one reported and explained by Micheli {\it et al.}~\cite{Micheli2010} and Qu{\'e}men{\'e}r  and Bohn.~\cite{Quemener2010b,Quemener2011} The other cases are new.   A useful figure of merit is a rate constant of $10^{-7}$ cm$^2/$s, which corresponds to a 1s lifetime of a 2D gas with density $n=10^7$ cm$^{-2}$, comparable to the number achieved experimentally with $^{40}$K$^{87}$Rb.~\cite{Ni2010}  Smaller rate constants correspond to lifetimes of longer than 1 s at this density. As the dipole strength $d(F)$ increases, $K_j^\mathrm{ls}$ decreases below $10^{-7}$ cm$^2/$s when $d(F)$ is large enough. Consequently, we predict that stabilization of the heavier non-reactive species RbCs and KCs in their low vibrational levels should be possible at dipole strength of only 1 D or less, at electric field below 10 kV/cm (see Fig.~\ref{fig2}). Of course, RbCs and KCs in v=0 is nonreactive.   Even a light reactive species LiK with weaker confinement should be stabilized in such a trap at a field on the order of 20 kV/cm.

Figure~\ref{fig6} also demonstrates the difference between the bosonic and fermionic form of the same species LiK and KCs.  As already shown for KRb, the fermionic rate is suppressed at zero field relative to the bosonic rate, due to the centrifugal barrier to the $p$-wave.    At  200nK the suppression ratio $K_1^\mathrm{ls}/K_0^\mathrm{ls} \approx$ 10 and 100 for KCs and LiK respectively, similar to the ratios in 3D (see Fig.~\ref{fig3}).    As field increases, the fermionic and bosonic rate constants for the same species move away from the van der Waals limit and are both suppressed, eventually becoming nearly equal as $d(F)$ continues to increase.  The figure also shows a comparison to an approximate formula which applies to bosons and fermions in the strong dipole limit where $a_d  \gg  a_h$,
\begin{equation}
\label{eq:KSC}
  K_\mathrm{D}^\mathrm{ls} = 2\frac{\hbar}{\mu} 288 (\kappa a_h)^4 e^{-3.0314\, (a_d/a_h)^{2/5}}  \,,
\end{equation}
where the factor 288 comes from fitting our numerical calculations.  The factor $2^{8/5}\approx 3.0314$ in the exponent is replaced by 2 if $a_d$ and $a_h$ are defined with the total molecular mass instead of the reduced mass.  The agreement with the calculations shows that this expression accurately captures the scaling with mass and $d(F)$ for a wide range of species.  It seems especially good for the bosons even when $a_d$ is only slightly larger than $a_h$.  The quantity to the right of $2\hbar/\mu$ in Eq.~(\ref{eq:KSC}) represents the transmission function $P_0(\Omega,F,E)$ and is similar to expressions for the semiclassical tunneling probability through the long range potential barrier proportional to $\exp\left (-\mathrm{C}\, (a_d/a_h)^{2/5}\right ) $, where $\mathrm{C}$ is a constant, as described by several authors.~\cite{Buchler2007,Micheli2007,Micheli2010,Ticknor2009,Ticknor2010,DIncao2011}  Since the rate constants $K_0^\mathrm{ls}$ and $K_1^\mathrm{ls}$ are nearly equal in the strong dipole limit, this implies $P_1\approx P_0/2$.   Our fit  for $P_0(\Omega,F,E) $ in Eq.~(\ref{eq:KSC}) is based on the calculated universal quantum dynamics  and does not rely on any semiclassical or adiabatic approximations.  It gives realistic scaling with the experimentally variable parameters:  trap strength $\Omega$,  electric field $F$ and temperature, which respectively control  confinement $a_h$,  dipole length $a_d$ and the range of collision energy $E$.

\begin{figure}[h]
\centering
  \includegraphics[height=7cm]{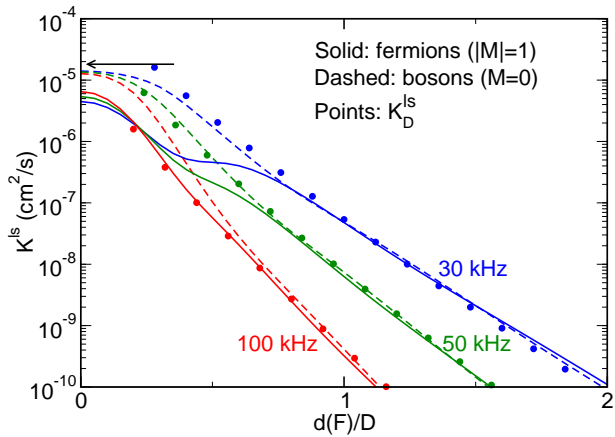}
  \caption{Universal loss rate constant $K_j^\mathrm{ls}$ for quasi-2D collisions of bosonic $^7$Li$^{133}$Cs (dashed lines) and fermionic $^6$Li$^{133}$Cs (solid lines) at $E/k_B=200$ nK for a range of trap strengths $\Omega/(2\pi)$.  The solid points show the fit $K_D^\mathrm{ls}$ in Eq.~\ref{eq:KSC} when $a_d > a_h$.  The arrow indicates the unitarity limit for bosons, $2\hbar/\mu$.  }
  \label{fig7}
\end{figure}

\begin{figure}[h]
\centering
  \includegraphics[height=7cm]{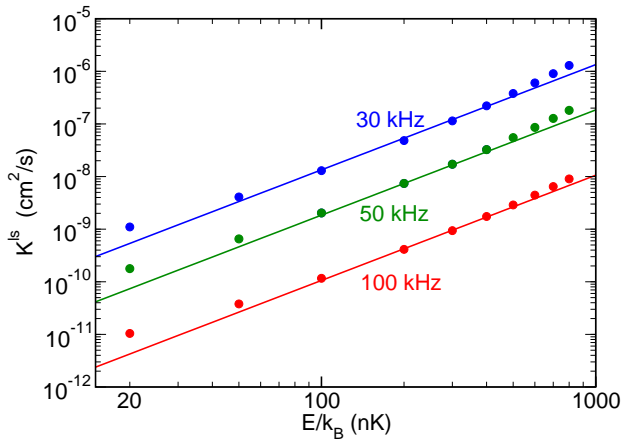}
  \caption{Universal loss rate constant $K_0^\mathrm{ls}$ (points) versus collision energy $E/k_B$ for  bosonic $^7$Li$^{133}$Cs at a dipole strength of $d=0.4$ au (1 D) for $\Omega/(2\pi)=$ 30 kHz, 50 kHz, and 100 kHz. The solid line shows the fit function in Eq.~(\ref{eq:KSC}).}
  \label{fig8}
\end{figure}

Figure~\ref{fig7} shows our calculations for bosonic and fermionic isotopes of the reactive species LiCs, which has the largest dipole moment of any of the mixed alkali species.  Since the dipole moment varies slowly with vibrational quantum number, this rate should apply to a range of low vibrational levels.  Stabilization in quasi-2D geometry should be possible at relatively modest electric fields.  The formula in Eq.~(\ref{eq:KSC}) is again very good for this species, verifying that in the strong dipole limit the probability $P_j$ decreases with increasing $\Omega$.  This is unlike the van der Waals limit, where $P_j$ from Eq.~(\ref{eq:KvdW}) scales as $\sqrt{\Omega}$.  Consequently Fig.~\ref{fig7} shows the switching between the two limits for dipole strengths where $a_d \approx a_h$.  Again, we find that the universal loss rates for bosons and fermions are nearly the same in the strong dipole limit.

Figure~\ref{fig8} shows that the $K_j^\mathrm{ls}$ in the strong dipole limit scales as $E^2$ for $j=0$ and 1 in the interesting experimental range of  50 nK to 500 nK, consistent with the $E^2$ scaling found by Ticknor.~\cite{Ticknor2010}  Some departure from this scaling is evident below 50 nK and above 500 nK in Fig.~\ref{fig8}.  In contrast, Eq.~(\ref{eq:KvdW}) shows that in the van der Waals limit as $F \to 0$ $K_1^\mathrm{ls}$ scales as $E$ and $K_0^\mathrm{ls}$ is nearly independent of $E$.

We have concentrated here on reactive or loss collisions in quasi-2D geometry, instead of elastic collisions.  We have also calculated the latter as a function of field.    Micheli {\it et al}~\cite{Micheli2010} found that a simple unitarized Born approximation was accurate for the two lowest partial waves for ultracold KRb. As dipole increases it is necessary to include more than the lowest partial wave.  Simple universal expressions for the strong dipole limit can be worked out.~\cite{Ticknor2009,DIncao2011} The only important aspect of quasi-2D elastic scattering to note here is that the elastic rate constant continues to increase towards the unitarity limit (of several partial waves) as electric field increases, and thus the ratio of elastic to loss collison rates becomes very large with increasing field, due to the strong suppression of the latter.  Thus, it is hoped that fast elastic collisions can thermalize a quasi-2D gas of molecular dipoles, which are stable with long lifetimes relative to loss on the time scale of an experiment.

\section{Conclusion}

We have characterized the universal reactive and inelastic relaxation rate constants in free space and in quasi-2D planar geometry of the ten mixed alkali-metal dimer polar molecular species in the near-threshold limit of ultracold collisions where the lowest partial waves contribute.   We consider collisions between molecules in a single state of vibration, rotation ($J=0$) and internal spin, where the bosonic or fermionic character of the molecule leads to different rate constants.  The universal rate constants are independent of the details of short range chemical interactions and depends only on the long range approach of the colliding species, which is subject to experimental control by electric and electromagnetic forces.   The universal rates apply as long as short range chemical dynamics has unit probability of chemical reaction or inelastic quenching that results in loss of the initially prepared ultracold molecules.  The species LiNa, LiK, LiRb, LiCs, and KRb have such unit  short range chemical reaction probabilities when in their vibrational ground state $v=0$, and are expected to have universal rate constants in all $v$.  The species NaK, NaRb, NaCs, KCs and RbCs are non-reactive in $v=0$, but may have universal vibrational quenching rate constants when in vibrationally excited states with $v \ge 1$.    The non-universal collisions of the $v=0$ level of these non-reactive species are predicted to have electric field-dependent scattering resonances,~\cite{Ticknor2005,Roudnev2009,Idziaszek2010b,DIncao2011} which remain to be more fully explored theoretically and experimentally.

In the absence of an electric field, the universal van der Waals rate constants are determined only by the van der Waals $C_6$ coefficient, the reduced mass of the pair, and the collision energy according to the known threshold laws.  The important parameter is the van der Waals length $\bar{a}$, which scales as $(\mu C_6)^{1/4}$ and thus is relatively insensitive to $C_6$.  We use simple estimates of $C_6$ based on the electronic and rotational properties of the molecules at their equilibrium internuclear separation $R_e$ that should be adequate approximations for low values of vibrational quantum number $v$.    Although the normal second-order expansion of the dispersion energy breaks down when two molecules approach one another when the dipolar interaction exceeds the rotational spacing, the distance at which this happens is around half of $\bar{a}$, where it does not significantly affect the reaction rate due to quantum scattering at distances on the order of $\bar{a}$ or larger.   We find that the ultracold universal reactive or inelastic loss rate constants for bosonic mixed alkali-metal dimers should be within a few factors of the unitarity upper bound to the rate constant, whereas rate constants for fermionic species tend to be significantly smaller by factors of 10 to 100 around 200 nK, scaling linearly with temperature.

When an electric field $F$ is present, the collision rates vary with $F$ and switch from the van der Waals universal limit to a universal strong dipolar limit.  We show that the universal ultracold collision rates are not significantly affected by the F-dependent changes in the second-order dispersion interaction, but can be calculated in 3D from rotationally adiabatic potentials.  We consider in detail the $F$-dependent universal  loss rate constants in quasi-2D geometry when the molecules are oriented in the $z$-direction by an electric field and tightly confined in that direction by a harmonic trap such that $\kappa a_h \ll 1$, where $\hbar \kappa$ is the collisional momentum and $a_h$ is the harmonic confinement length.  Our fully numerical calculations connect the wave function in the universal  3D short range core of the collision with incoming boundary conditions to the reduced dimensional quasi-2D wave function at long range.  As in 3D the collision switches from the van der Waals limit to a strong dipole limit as $F$ increases, where the rate constants for bosons and fermions of the same species are the same.  We obtain a simple scaling formula that provides a good approximation of the universal rate constants for all ten species in the ultracold domain below 1 $\mu$K and for trapping frequencies in the tens of kHz range that are experimentally feasible.  The universal strong dipolar limit of the rate constant has simple scaling with the mass, trapping frequency, and molecular dipole strength.  In this geometry the dipoles have long range repulsive interactions, and universal collision rate constants can be decreased to be several orders of magnitude less than the unitarity limit, so that gases of oriented dipoles of even highly reactive species can be stabilized against loss collisions under practical experimental conditions that should be achievable with current techniques for cooling, trapping, and optical lattice design and with relatively modest electric fields on the order of 10 kV/cm .  We expect these considerations will also be valid for quasi-1D geometry, or collisions in ``tubes'' with tight confinement in the two directions orthogonal to the ``tube.''

While we have considered molecular collisions for states with low vibrational quantum numbers $v$, the van der Waals $C_6$ coefficient and, perhaps more importantly, the molecular dipole moment will vary with $v$.  Consequently, there is some degree of control possible by ``vibrational tuning'' of these parameters.  Eventually, at very high $v$ near the dissociation limit, the expectation value of the dipole moment will become very small, and electric field control will be lost.  But there is a wide range of vibrational levels that should be accessible for experimental control, so that the $v=0$ level need not be the target level of STIRAP, especially if higher $v$ levels are easier to reach.  Since the short range collision loss probability is maximal for universal collisions, it is not necessary to have a non-reactive $v=0$ species to work with, if collisions can be shielded in quasi-2D geometry.    One could envision achieving a lattice of ``preformed pairs'' of unassociated atoms in an array of fully confining 3D trapping cells (zero-dimensional collisions of atoms bound to the cell), as has been proposed.~\cite{Damski2003,Freericks2010} Converting such atom pairs to Feshbach molecules by magnetoassociation and moving the latter to a desired $v$ by STIRAP should allow the study of controlled collisions upon subsequent removal of the fully confining lattice by turning off some or all of the lattice lasers, resulting in lattices with quasi-1D or quasi-2D geometry or even a free space gas.  A broad class of such collisions should have rate constants approximated by the universal ones described here.  

\section{Acknowledgments}

This work was supported in part by an AFOSR MURI grant on ultracold polar molecules and in part by a Polish Government Research Grant for the years 2011-2014.

%The conclusions section should come at the end of article. For the reference section, the style file rsc.bst can be used to generate the correct reference style

%\balance


\footnotesize{

%\bibliography{../BibFiles/Allrefs} %your .bib file
\bibliography{References} %your .bib file

\bibliographystyle{rsc} %the RSC's .bst file
}

\end{document}